\documentstyle[aas2pp4,psfig]{article}
\newcommand\mdot   {\hbox {${\dot M}$}}
\newcommand\sz     {$S_{\rm z}$}
\newcommand\mzon   {M$_{\odot}$}
\newcommand\pp     {$\pm$}

\newcommand\micros  {$\mu$s}

\righthead{KiloHertz Quasi-Periodic Oscillations in GX\,5$-$1}
\slugcomment{Submitted to ApJ Letters, June 1998}

\begin{document}

\title{Discovery of KiloHertz Quasi-Periodic Oscillations in the Z
source GX\,5$-$1}

\author{Rudy Wijnands\altaffilmark{1},
        Mariano M\'endez\altaffilmark{1,2},	
        Michiel van der Klis\altaffilmark{1,3},
	Dimitrios Psaltis\altaffilmark{4},
	Erik Kuulkers \altaffilmark{5},
	Frederick K. Lamb\altaffilmark{6}
        }

\altaffiltext{1}{Astronomical Institute ``Anton Pannekoek'',
University of Amsterdam, and Center for High Energy Astrophysics,
Kruislaan 403, NL-1098 SJ Amsterdam, The Netherlands;
rudy@astro.uva.nl, mariano@astro.uva.nl michiel@astro.uva.nl}

\altaffiltext{2}{Facultad de Ciencias Astron\'omicas y
Geof\'{\i}sicas, Universidad Nacional de La Plata, Paseo del
Bosque S/N, 1900 La Plata, Argentina}

\altaffiltext{3}{Department of Astronomy, University of California,
Berkeley, Berkeley, CA 94720}

\altaffiltext{4}{Harvard-Smithsonian Center for Astrophysics, 60
Garden St., Cambridge, MA 02138; dpsaltis@cfa.harvard.edu}

\altaffiltext{5}{Astrophysics, University of Oxford,
Nuclear and Astrophysics Laboratory, Keble Road, Oxford OX1 3RH,
United Kingdom; e.kuulkers1@physics.oxford.ac.uk}

\altaffiltext{6}{Departments of Physics and Astronomy,
University of Illinois at Urbana-Champaign, Urbana, IL 61801;
f-lamb@uiuc.edu}

\begin{abstract}
We discovered two simultaneous kHz quasi-periodic oscillations (QPOs)
in the bright low-mass X-ray binary and Z source GX 5$-$1 with the
{\it Rossi X-ray Timing Explorer}. In the X-ray color-color and
hardness-intensity diagram a clear Z track is traced out, which
shifted between observations. The frequencies of the two kHz QPOs
increased from $\sim$215 Hz and $\sim$500 Hz on the left part of the
horizontal branch to $\sim$700 Hz and $\sim$ 890 Hz, respectively, on
the upper part of the normal branch. With increasing frequency the
FWHM and rms amplitude (8.6--60 keV) of the higher-frequency kHz QPO
decreased from 300 to 30 Hz , and from 6.6 to 2.4\%, respectively. The
FWHM and amplitude of the lower-frequency kHz QPO (50--100 Hz and
3--4\% rms) did not correlate with the position of the source on the Z
track. The kHz QPO separation was consistent with being constant at
298\pp11 Hz.  Simultaneously with the kHz QPOs horizontal branch
oscillations (HBOs) were detected with frequencies between 18 and 56
Hz.

\end{abstract}


\section{Introduction \label{intro}}

The bright low-mass X-ray binary (LMXB) and Z source (Hasinger \& van
der Klis 1989) GX 5$-$1 traces out a clear Z track in the X-ray
color-color diagram (CD). The limbs of this track are called the
horizontal branch (HB), normal branch (NB), and flaring branch
(FB). It is thought (e.g., Hasinger \& van der Klis 1989) that the
mass accretion rate (\mdot) increases when the source moves from the
left end to the right end of the HB, down the NB, and onto the
FB. Two types of quasi-periodic oscillations (QPOs) were so far
detected in GX 5$-$1 : on the HB oscillations called HBO were detected
with a frequency between 13 and 50 Hz (van der Klis et al. 1985), and
on the NB oscillations called NBO with a frequency of $\sim$6 Hz
(e.g., Lewin et al. 1992). No QPOs were detected on the FB of
GX\,5$-$1.

Recent, two simultaneously kHz QPOs were discovered in the Z sources
Sco\,X-1 (van der Klis 1996a; 1997), GX 17$+$2 (Wijnands et
al. 1997b), Cyg X-2 (Wijnands et al. 1998), GX 340$+$0 (Jonker et
al. 1998), and GX 349$+$2 (Zhang, Strohmayer, \& Swank 1998). The
properties of these QPOs are remarkably similar to each other. They
are best detected on the HB and the upper part of the NB. So far, Sco
X-1 is the only Z sources in which kHz QPOs have been detected on the
lower part of the NB and the FB. The frequencies of the kHz QPOs
increase from the left end of the HB to the upper part of the NB. In
Sco X-1, their frequencies continue to increase as the source moves
down the NB and onto the lower part of the FB.  Similar kHz QPOs have
been detected (e.g. Strohmayer et al. 1996; Berger et al. 1996; Zhang
et al. 1996) in the less luminous LMXBs, called the atoll sources
(Hasinger \& van der Klis 1989). The many similarities (e.g., the
frequencies and the dependence of the kHz QPO parameters on inferred
\mdot) between the kHz QPOs detected in the Z and atoll sources
suggest that these QPOs are produced by the same physical mechanism in
both types of LMXBs (see van der Klis 1998 for a review of kHz QPOs).

In this paper we report the discovery of two simultaneous kHz QPOs in
the Z source GX 5$-$1.  A preliminary announcement of the discovery of
the kHz QPO was already made by van der Klis et al. (1996b).

\section{Observations and Analysis  \label{observations}}

We observed GX 5$-$1 on 1996 November 2, 6, and 16 (the AO1 data), and
on 1997 May 30, June 5, and July 25 (the AO2 data) using the
proportional counter array onboard the {\it Rossi X-ray Timing
Explorer} (RXTE). We obtained a total of 102 ksec of good data. During
17 \% of the observing time only 3 or 4 out of the 5 detectors were
active. In constructing the CD and the hardness-intensity diagrams
(HIDs) we used only the data of the three detectors that were always
on. For the power spectral analysis we used all available data. Data
were obtained in 129 photon energy channels (2--60 keV) with a time
resolution of 16 s. Simultaneous data were collected in two photon
energy bands (2--8.6 and 8.6--60 keV; the AO1 data) or four (2--5.0,
5.0--6.4, 6.4--8.6, and 8.6--60 keV; the AO2 data) with a time
resolution of 122 \micros.  The CD and HIDs were constructed using the
16 s data; power density spectra were calculated using the 122
\micros\, data.

To determine the properties of the kHz QPOs, we fitted the 64--4096 Hz
power density spectra with a function containing one or two
Lorentzians (to represent the kHz QPOs), a constant plus a broad
sinusoid (the dead-time modified Poisson level; Zhang et al. 1995),
and sometimes an extra Lorentzian with a centroid frequency below 50
Hz to represent the power spectral continuum below 200 Hz. The Very
Large Event (van der Klis et al. 1997) window was set to 55
\micros. Its effect on the Poisson noise was small and could be
incorporated in the broad sinusoid. The properties of the HBOs and the
NBOs were determined by fitting the 0.125--192 Hz power density
spectra with one or two Lorentzians (the HBO and its second harmonic;
the NBO), and an extra Lorentzian (with a centroid frequency $\sim$0
Hz) to represent the underlying continuum.

The uncertainties in the fit parameters were determined using
$\Delta\chi^2=1$ and upper limits using $\Delta\chi^2=2.71$,
corresponding to 95\% confidence levels. Upper limits on the kHz QPOs
amplitude were determined using a FWHM of 150 Hz. When only one of the
kHz QPOs was detected, the upper limit on the other kHz QPO was
determined by fixing its frequency to the frequency of the detected
kHz QPO plus or minus (depending on which kHz QPO was detected) the
mean peak separation. Upper limits on the amplitude of the HBO and its
second harmonic were determined using a FWHM of 10 Hz or 20 Hz,
respectively, and the frequency of the second harmonic was fixed at
twice the frequency of the HBO.

In the CD and the HIDs, the soft color is defined as the logarithm of
the 3.5--6.4\,keV/2.0--3.5\,keV count rate ratio, the hard color as
the logarithm of the 9.7--16.0\,keV/6.4--9.7\,keV count rate ratio,
and the intensity is the logarithm of the 2.0--16.0 keV count rate. By
using logarithmic values for the colors and the intensity, the \sz\,
parameterization (see Section \ref{results}) does not depend on the
values of those quantities but only on their variations (Wijnands et
al. 1997a).  The count rates used in the diagrams are
background-subtracted but not dead-time corrected. The dead-time
correction is 3--5 \%.

\section{Results \label{results}}

\begin{figure}[t]
\begin{center}
\begin{tabular}{c}
\psfig{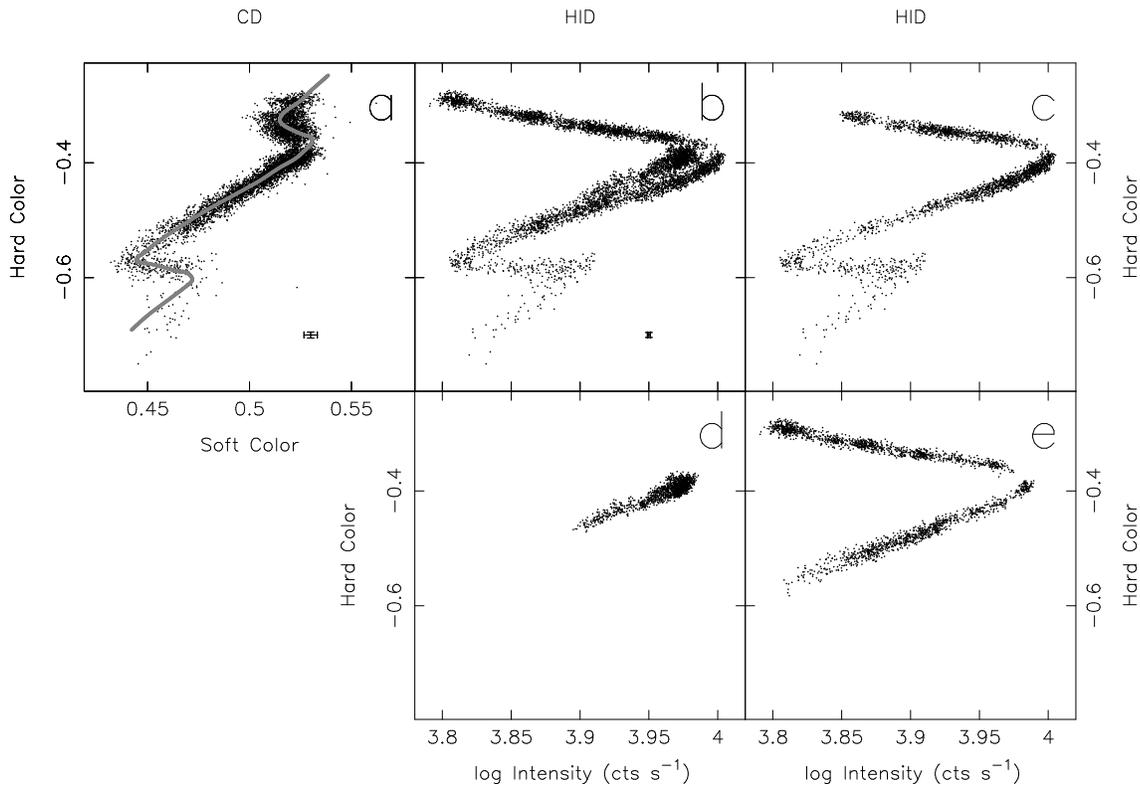}
\end{tabular}
\caption[]{The color-color and hardness-intensity diagrams of GX
5$-$1. (a) the CD of all data, (b) the HID of all data, (c) the HID of
the AO1 data, (d) the HID of the AO2 1997 May 30 and June 5 data, and
(e) the HID of the AO2 1997 July 25 data.  In all diagrams the soft
color is the logarithm of the 3.5--6.4 keV/2.0--3.5 keV count rate
ratio, the hard color the logarithm of the 9.7--16.0 keV/6.4--9.7 keV
count rate ratio, and the horizontal axis the logarithm of the
2.0--16.0 keV count rate. Background was subtracted but no deadtime
correction (3--5\%) has been applied. All points are 16 second
averages. Typical error bars are plotted in the bottom right corner of
(a) and (b). The gray solid line in (a) is the track used to
calculated the \sz\, values (see Section \ref{results}).
\label{CD_HID}}
\end{center}
\end{figure}

The CD and HID of all data combined is shown in Figures \ref{CD_HID}a
and b, respectively.  Clearly, in the HID (Fig. \ref{CD_HID}b) the Z
track is multiple. The reason for this is that it shifted slightly
between observations. The AO1 data form one Z track with an extra
branch trailing off the FB (Fig. \ref{CD_HID}c).  The AO2 data are
located at three different places in the HID. The 1997 July 25 data
(Fig. \ref{CD_HID}e) show a HB and NB which are displaced from
the AO1 track to lower count rates. The upper NBs of the AO1 and AO2
data do not fall on top of each other, but the lower NBs {\it do}
(Fig. \ref{CD_HID}b), indicating that the NB/FB vertex moved around
less than the HB/NB vertex. The 1997 May 30 and June 5 data cover a
segment of the NB (Fig. \ref{CD_HID}d) located slightly to the right
of the 1997 July 25 NB. The rapid X-ray variability (see below)
confirms that GX 5$-$1 was on the upper part of the NB during the 1997
May 30 and June 5 observations.  In the CD (Fig. \ref{CD_HID}a) the Z
track is clearly visible with the extra branch trailing the FB and a
slight upturn of the HB, which is only marginally visible in the HIDs
(Fig. \ref{CD_HID}b).  The motion of the Z track is in the CD obscured
by the errors in the X-ray colors.

Because of the motion of the Z track in the HID, we decided to analyse
the power spectra of the AO1, the May/June AO2 and the July AO2 data
(hereafter referred to as the AO2 data, unless otherwise mentioned)
separately.  We selected the power density spectra according to the
position of GX 5$-$1 on the Z track in the HIDs of the AO1
(Fig. \ref{CD_HID}c) and AO2 (Fig. \ref{CD_HID}e) data, and
measured for each selection the average position of GX 5--1 along the
Z track using the \sz\, parameterization (see e.g. Wijnands et
al. 1997a) applied to the combined CD (Fig. \ref{CD_HID}a).  The
errors on the \sz\, values are the standard deviation of each
selection.

\begin{figure}[t]
\begin{center}
\begin{tabular}{c}
\psfig{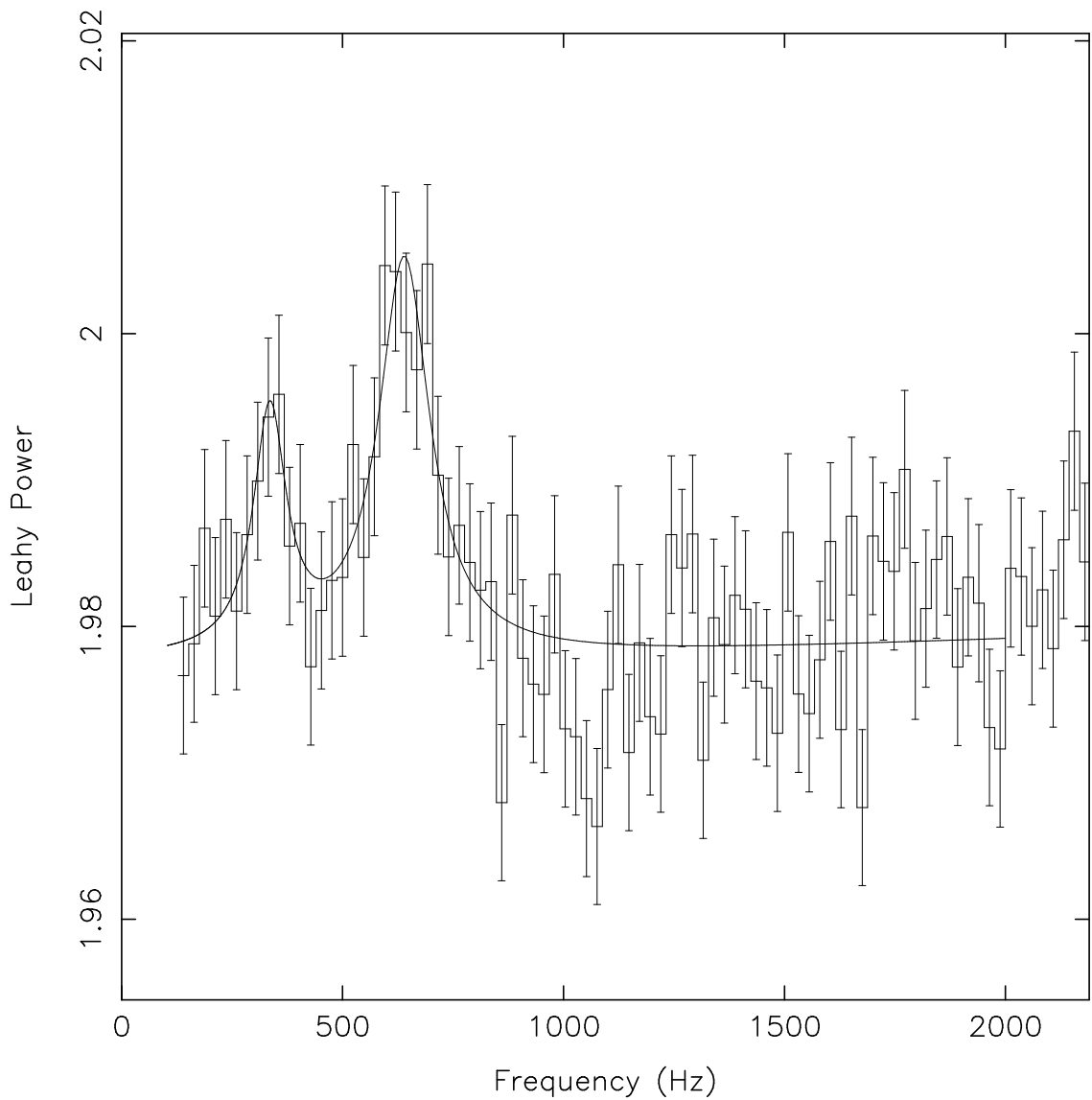}
\end{tabular}
\caption[]{Typical Leahy normalized power density spectrum in the energy
range 8.6--60 keV showing the two simultaneous kHz QPOs.
\label{powerspectra}}
\end{center}
\end{figure}

Two simultaneous kHz QPOs were detected in the AO1 and AO2 HB and
upper NB data (Fig. \ref{powerspectra}).  These kHz QPOs were most
significant in the highest photon energy band (8.6--60 keV). As adding
the lower energy data did not improve the significance, we used the
8.6--60 keV energy band throughout our analysis. The rms amplitude
upper limits and the detected values of the kHz QPOs at different
photon energies are displayed in Table \ref{energy}.  Note, that not
all kHz QPO detections were highly significant. For example, the three
detections of the kHz QPOs in the AO1 data with the highest \sz\,
value (see Fig. \ref{qpoversussz}a) are only significant at the
1.7--2.3 $\sigma$ level.

The properties of the kHz QPOs versus \sz\, are shown in Figure
\ref{qpoversussz}.  The two simultaneous kHz QPO peaks increased in
frequency (Fig. \ref{qpoversussz}a) from the left end of the HB
($\sim$500 Hz and $\sim$215 Hz, respectively) to the HB/NB vertex and
the upper part of the NB ($\sim$890 Hz and $\sim$700 Hz). Over this
interval the rms amplitude (Fig. \ref{qpoversussz}c) and the FWHM of
the higher-frequency peak decreased from 6.6\% to 2.4\%, and from 300
Hz to 30 Hz, respectively. The rms amplitude (3--4 \%;
Fig. \ref{qpoversussz}e) and FWHM (50--100 Hz) of the lower-frequency
kHz QPO show no clear relation with \sz. In the combined AO2 1997 May
30 and June 5 data kHz QPOs were detected only on the upper part of
the NB, with an rms amplitude, FWHM, and frequency of 2.0\pp0.4\%,
93$^{+47}_{-39}$Hz, and 557$^{+21}_{24}$Hz for the lower-frequency kHz
QPO, and 2.4$^{+0.5}_{-0.4}$\%, 122$^{+71}_{-43}$Hz, and
856$^{+19}_{-17}$Hz for the higher-frequency kHz QPO,
respectively. Lower down the NB, the upper limits on the kHz QPOs were
2.2\% rms and 3.4\% rms for the lower- and higher-frequency kHz QPOs,
respectively.  During the AO1, the AO2 May/June and the AO2 July data
the peak separation was consistent with being constant at 297\pp17 Hz,
298\pp11 Hz, and 299\pp15 Hz, respectively.  The average peak
separation for all observations combined was 298\pp11 Hz.

Simultaneously with the kHz QPOs, the HBO and its second harmonic were
detected. The frequency of the HBO increased from 18 to 56 Hz as the
source moved from the left end of the HB to the HB/NB vertex and onto
the upper part of the NB (Fig. \ref{qpoversussz}b). Over this interval
the rms amplitude of the HBO fundamental decreased from 11\% to 3\%
(8.6--60 keV; Fig. \ref{qpoversussz}d). The rms amplitude of the
second harmonic (detected only below \sz=1) decreased from 8\% to 3\%
(Fig. \ref{qpoversussz}f).  The FWHM of the fundamental increased from
5 to 24 Hz, and the FWHM of the second harmonic remained approximately
constant at $\sim$25 Hz. In the AO2 May/June data the HBO was also
detected simultaneously with the kHz QPOs, with an amplitude, FWHM,
and frequency of 4.70\pp0.9\% rms, 16.0\pp0.8 Hz, and 50.9\pp0.2 Hz,
respectively.  NBOs with frequencies of $\sim$5.7 Hz were detected
between \sz$\sim$1.1 and \sz$\sim$1.8. The NBOs were detected
simultaneously with the HBO fundamental on the upper NB, but never
simultaneously with the HBO second harmonic. No QPOs were detected
further down the NB, on the FB, and on the extra branch.

\section{Discussion \label{discussion}}

We have discovered two simultaneous kHz QPOs on the HB of the Z source
GX\,5$-$1. KHz QPOs have now been discovered in all six of the known Z
sources.  The frequencies of both peaks increased with inferred
\mdot. The rms amplitude and the FWHM of the higher-frequency kHz QPO
decreased as its frequency increased; the rms and the FWHM of the
lower-frequency kHz QPO remained roughly constant.  In Sco X-1 the
peak separation decreased when the source moved from the upper NB to
the lower NB onto the FB. The peak separation for GX 5$-$1 is
consistent with being constant, but also with being similar to
Sco\,X-1. This is true for the other Z sources as well, so that in the
simplest description {\it all} peak separations vary, providing a
challenge for beat frequency models where the peak separation is
identified with the neutron star spin rate (see Psaltis et al. 1998
for a critical discussion of the peak separations).

In the leftmost part of the HB, i.e. at the lowest inferred \mdot,
during the AO2 observation we detected the lower-frequency kHz QPO at
a frequency of 213.6$^{+11.1}_{-7.5}$ Hz.  The corresponding
higher-frequency kHz QPO has also a rather low frequency
(504.5$^{+10.4}_{-12.6}$ Hz), but not unusually low (see
e.g. Strohmayer et al. 1996; Ford et al. 1997; Jonker et al. 1998).
The lower-frequency kHz QPO covers a total range in frequency of a
factor 3.5. This QPO has the lowest frequency and the largest dynamic
range for a kHz QPO so far detected in any LMXB. The frequency range
falls in the 200--1000 Hz frequency range expected if the
lower-frequency kHz QPO is identified with the beat of the Keplerian
frequency at the sonic-point with the neutron star spin frequency
(Miller, Lamb, \& Psaltis 1998). The corresponding range for the
sonic-point radius would be 17--26 km, assuming a neutron star mass of
1.4 \mzon\, and that the peak separation is the spin frequency.

Recently, Stella \& Vietri (1998) proposed that the low frequency QPOs
observed in atoll sources are due to a precession of the innermost
disk region, dominated by the Lense-Thirring effect. They also
suggested that the HBO in Z sources could arise from the same
mechanism. However, the precession frequencies predicted for GX 5$-$1
are about a factor 2.5 smaller than that of the HBO frequencies (see
also Stella \& Vietri 1998), unless I/M (with I the moment of inertia
of the neutron star and M its mass) is a factor of 2.5 higher than
predicted by neutron star models with realistic equations of state,
the kHz QPO peak difference is half the neutron star spin frequency,
or the HBOs are not the fundamental frequencies but the second
harmonics. For a detailed discussion about the Lense-Thirring
interpretation of the HBO in Z sources we refer to Psaltis et
al. (1998)

\begin{figure}[t]
\begin{center}
\begin{tabular}{c}
\psfig{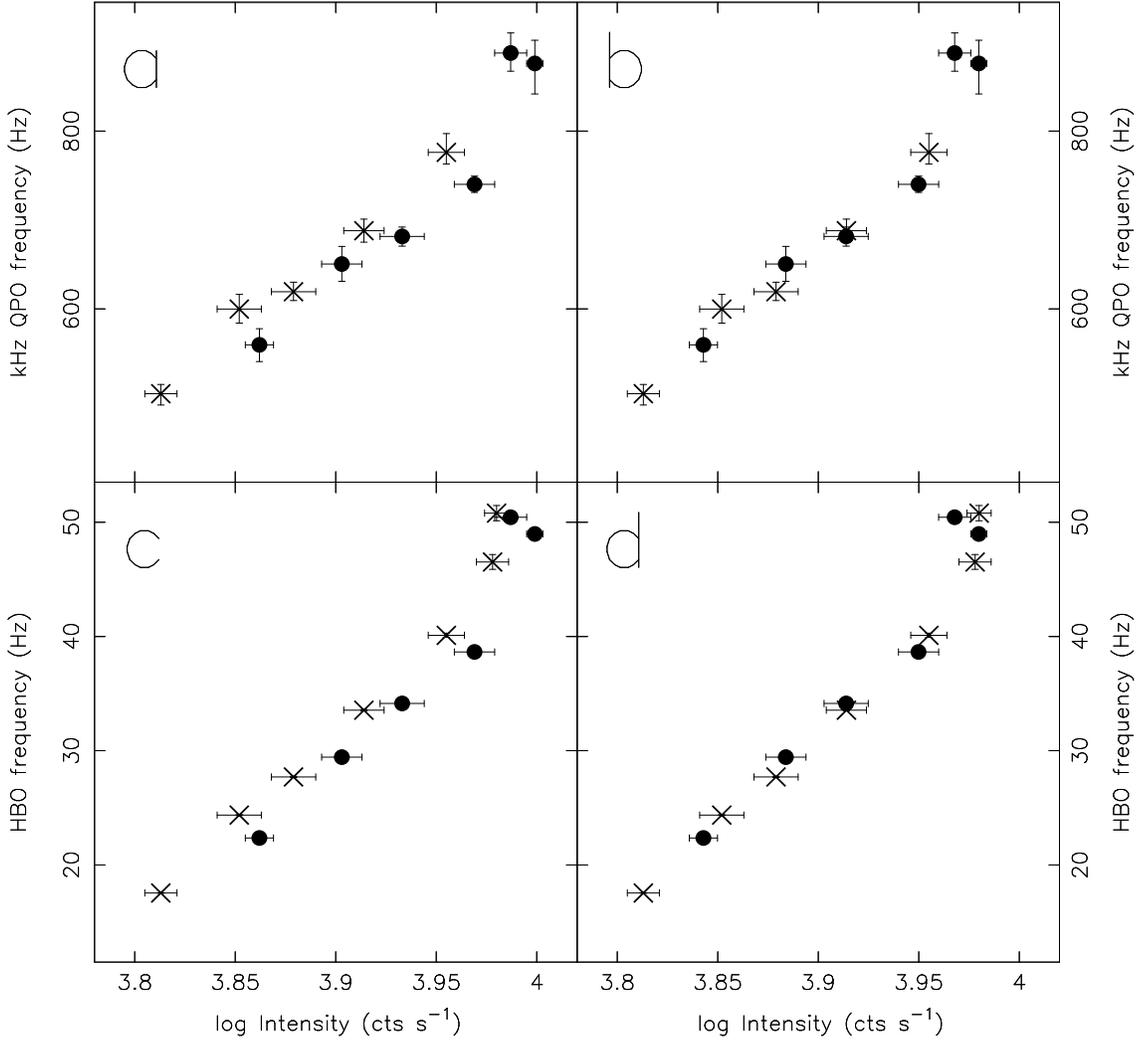}
\end{tabular}
\caption[]{The higher-frequency kHz QPO frequencies ({\it a} and {\it
b}) and the HBO frequencies ({\it c} and {\it d}) versus the
intensity. The definition of the intensity is the same as in
Fig. \ref{CD_HID}.  The bullets are the AO1 data and the crosses the
AO2 July 25 data. In {\it b} and {\it d} the count rates of the AO1
data were corrected for the shift of the Z track (see
Fig. \ref{CD_HID}; Section \ref{discussion}).
\label{qpocr}}
\end{center}
\end{figure}

The Z track of GX 5$-$1 shifts between observations. The Z track of
the AO1 data is at $\sim$4.5\% higher count rates and $\sim$2\% softer
hard colors than the Z track of the AO2 data. Similar shift were
already reported by Kuulkers et al. (1994) using EXOSAT data, although
in their observations the variations in count rate and colors were
slightly larger ($\sim$8\% and $\sim$6\%, respectively).  When
plotting the frequency of the higher-frequency kHz QPO and the HBO
versus count rate (Fig. \ref{qpocr}a and c) the AO1 data are slightly
displaced to higher count rates compared to the AO2 data.  If we
correct the count rates of the AO1 data for the shifts in count rate
derived above the kHz QPOs and HBO frequencies as a function of count
rate are consistent with being the same during both epochs
(Fig. \ref{qpocr}b and d).  The same is also true for the frequency of
the lower-frequency kHz QPO (not shown here) and the rms and the FWHM
of all the QPOs.  The only difference could be the rms amplitude of the
higher-frequency kHz QPOs at the lowest inferred mass accretion rate,
however, the difference is marginal. This demonstrates that not the
X-ray count rate determines the properties of the QPOs (both the kHz
QPOs {\it and} the HBO), but the position of GX 5$-$1 on the Z track
does. Kuulkers et al. (1994) also reported that the HBO properties
were consistent with being the same when the Z track of GX 5$-$1 was
displaced in the HID.  In their observations the tracks were displaced
more than our tracks indicating that even for larger shifts the HBO
properties do not change.  Jonker et al. (1998) found the same for the
HBO (the statistics did not allow to conclude anything for the kHz
QPOs) in GX 340$+$0.  For Cygnus X-2 (Wijnands et al. 1997) it was
shown that the NBO properties changed significantly when the Z track
of this source moved around in the HID. However, due to limited
available data nothing could be concluded for the HBO properties in
Cygnus X-2.

\acknowledgments

This work was supported in part by the Netherlands Foundation for
Research in Astronomy (ASTRON) grant 781-76-017 and by NSF grant AST
96-18524.  F.K.L acknowledges support from the United States National
Aeronautics and Space Administration (NAG 5-2925).  M.M. is a fellow
of the Consejo Nacional de Investigaciones Cient\'{\i}ficas y
T\'ecnicas de la Rep\'ublica Argentina. M.K. gratefully acknowledges
the Visiting Miller Professor Program of the Miller Institute for
Basic Research in Science (UCB).

\clearpage

\begin{deluxetable}{ccccc}
\tablecolumns{5}
\footnotesize
\tablewidth{0pt}
\tablecaption{Amplitude of the kHz QPOs$^a$
\label{energy}}
\tablehead{
Energy  & Lower-frequency peak$^b$ & Higher-frequency peak$^c$\\
  (keV) &      (rms)  & (rms)}
\startdata
2--5.0   &  $<1.3$\% & $<2.0$\% \\
5.0--6.4 &  $<2.2$\% & $<2.6$\% \\
6.4--8.6 &  $<2.3$\% & 2.6\pp0.5\%\\
8.6--60  &  3.1$^{+0.6}_{-0.4}$\% & 4.7\pp0.4\%\\
\enddata
\tablenotetext{a}{The values were calculated at \sz=0.86\pp0.05}
\tablenotetext{b}{FWHM = 98$^{+53}_{-34}$ Hz, frequency =
340$^{+11}_{-13}$ Hz}
\tablenotetext{c}{FWHM = 150$^{+28}_{-24}$ Hz, frequency = 647\pp10 Hz}
\end{deluxetable}

\clearpage

\clearpage

\clearpage

\begin{figure}[t]
\begin{center}
\begin{tabular}{c}
\psfig{figure=QPOs_AO1_AO2.ps}
\end{tabular}
\caption[]{The properties of the kHz QPOs versus postion on the Z track,
\sz\, (see Section \ref{results}); (a) frequency of the kHz QPOs, (b)
frequency of the HBO fundamental and 2nd harmonic, (c) the rms
amplitude of the higher-frequency kHz QPO, (d) rms amplitude of the
HBO fundamental, (e) rms amplitude of the lower-frequency kHz QPO, and
(f) rms amplitude of the HBO 2nd harmonic.  The black points are the
AO1 data and the red points the AO2 July 25 data.
\label{qpoversussz}}
\end{center}
\end{figure}
\clearpage

\end{document}